\documentclass[twocolumn]{aastex631}

\usepackage{ifthen}
\usepackage{afterpage}

\newcommand{\comment}[1]{}

\newcommand{\revision}[1]{{#1}}

\usepackage{amsmath}
\usepackage{natbib}
\usepackage{appendix}
\usepackage{gensymb}
\usepackage{wasysym}
\usepackage{bm}

\providecommand{\adsurl}[1]{\href{#1}{ADS}}

\newcommand{\lmax}{$l_{\mathrm{max}}$}

\DeclareSymbolFont{UPM}{U}{eur}{m}{n}
\DeclareMathSymbol{\umu}{0}{UPM}{"16}
\let\oldumu=\umu
\renewcommand\umu{\ifmmode\oldumu\else$\oldumu$\fi}
\newcommand\micro{\umu}

\newcommand\microns{\micro m}

\shorttitle{LOO-CV Eclipse Mapping}
\shortauthors{Challener, Welbanks, \& McGill}

\begin{document}

\title{Bringing 2D Eclipse Mapping out of the Shadows with Leave-one-out Cross-validation}
\author[0000-0002-8211-6538]{Ryan C. Challener}
\affiliation{Department of Astronomy, University of Michigan, 1085
  S. University Ave., Ann Arbor, MI 48109, USA}
\affiliation{Department of Astronomy, Cornell University, 122 Sciences Drive, Ithaca, NY 14853, USA}

\author[0000-0003-0156-4564]{Luis Welbanks}
\thanks{NHFP Sagan Fellow}
\affil{School of Earth \& Space Exploration, Arizona State University, Tempe, AZ, 85257, USA}

\author[0000-0002-1052-6749]{Peter McGill}
\affiliation{Department of Astronomy and Astrophysics, University of California, Santa Cruz, CA 93105, USA}
\affiliation{Lawrence Livermore National Laboratory, 7000 East Ave., Livermore, CA}



\begin{abstract}


Eclipse mapping is a technique for inferring 2D brightness maps of transiting exoplanets from the shape of an eclipse light curve.
With JWST's unmatched precision, eclipse mapping is now possible for a large number of exoplanets.
However, eclipse mapping has only been applied to two planets and the nuances of fitting eclipse maps are not yet fully understood.
Here, we use Leave-one-out Cross-Validation (LOO-CV) to investigate eclipse mapping, with application to a JWST NIRISS/SOSS observation of the ultra-hot Jupiter WASP-18b.
LOO-CV is a technique that provides insight into the out-of-sample predictive power of models on a data-point-by-data-point basis.
We show that constraints on planetary brightness patterns behave as expected, with large-scale variations driven by the phase-curve variation in the light curve and smaller-scale structures constrained by the eclipse ingress and egress.
For WASP-18b we show that the need for higher model complexity (smaller-scale features) is driven exclusively by the shape of the eclipse ingress and egress.
We use LOO-CV to investigate the relationship between planetary brightness map components when mapping under a positive-flux constraint to better understand the need for complex models.
Finally, we use LOO-CV to understand the degeneracy between the competing ``hotspot'' and ``plateau'' brightness map models of WASP-18b, showing that the plateau model is driven by the ingress shape and the hotspot model is driven by the egress shape, but preference for neither model is due to outliers or unmodeled signals.
Based on this analysis, we make recommendations for the use of LOO-CV in future eclipse-mapping studies.

\end{abstract}


\section{INTRODUCTION}

When an exoplanet passes behind its host star, the brightness pattern on the planet's atmosphere is imprinted on the shape of the eclipse light curve \citep{CowanFujii2018haexMappingReview}.
By inverting the light curve, we can recover a map of the planet's spatial brightness and, thus, temperature distribution.
Such maps will be crucial to understanding the fundamental processes of these atmospheres, like winds and magnetic fields, and will be critical to refining multidimensional aspects of general circulation models (GCMs).

To date, two planets have been mapped in eclipse: HD 189733 b \citep{DeWitEtal2012aaHD189Map, MajeauEtal2012apjlHD189Map, RauscherEtal2018ajMap, ChallenerRauscher2022ajThERESA} and WASP-18b \citep{CoulombeEtal2023arxivWASP18b}.
The map of HD 189733 b required six \textit{Spitzer} 8 \microns\ photometric eclipses and a partial phase curve to constrain the planet's eastward hotspot offset and dayside temperatures.
WASP-18b was mapped as part of the JWST Transiting Exoplanet Community Early Release Science Program (ID: 1366) using a single NIRISS/SOSS eclipse.
The light curve was fit equally well by a model with a dayside hotspot ``plateau'', which has a similar temperature across much of the dayside, and a model with a significantly non-equatorial hotspot.
Comparison between models using traditional model selection metrics like $\chi^2$ and the Bayesian Information Criterion \citep[BIC,][]{Raftery1995BIC} showed no preference for either model \citep{CoulombeEtal2023arxivWASP18b}.

These selection metrics provide a single value to assess a model fit to an entire data set, providing little insight into how the data inform the models. Recent advances in approximate Bayesian computation have enabled more interpretable cross-validation model selection and criticism scores to be computed \citep{vehtari2015pareto}. In particular, Bayesian Leave-One-Out Cross-Validation  \citep[LOO-CV;][]{vehtari2017practical} and has seen application to a variety of astronomical datasets \citep[e.g.,][]{Morris2021,Meier2022, Neil2022, McGill2023}. Recently, \cite{WelbanksEtal2023ajLOOCV} demonstrated that LOO-CV is a powerful tool for understanding modeling inference at the level of individual data points with application to exoplanet atmospheric retrieval. These inference techniques are traditionally thought of as data-driven statistical methodologies. However, a persistent problem when analyzing these data has been whether competing explanations are data- or model-driven. With LOO-CV, \cite{WelbanksEtal2023ajLOOCV} showed how the use of per-data-point scores can help understand which data points drive the detections in an atmospheric spectrum and their associated atmospheric constraints. 

In this work, we use LOO-CV to provide insight into eclipse mapping models. In Section \ref{sec:mapping} we describe our eclipse-mapping model, in Section \ref{sec:loocv} we describe LOO-CV, in Section \ref{sec:application} we apply LOO-CV to an eclipse-mapping analysis, and in Section \ref{sec:conclusions} we lay out our conclusions and recommendations for future studies. 

\section{Eclipse Mapping with Eigenmaps}

\label{sec:mapping}

\begin{figure*}
    \centering
    \includegraphics{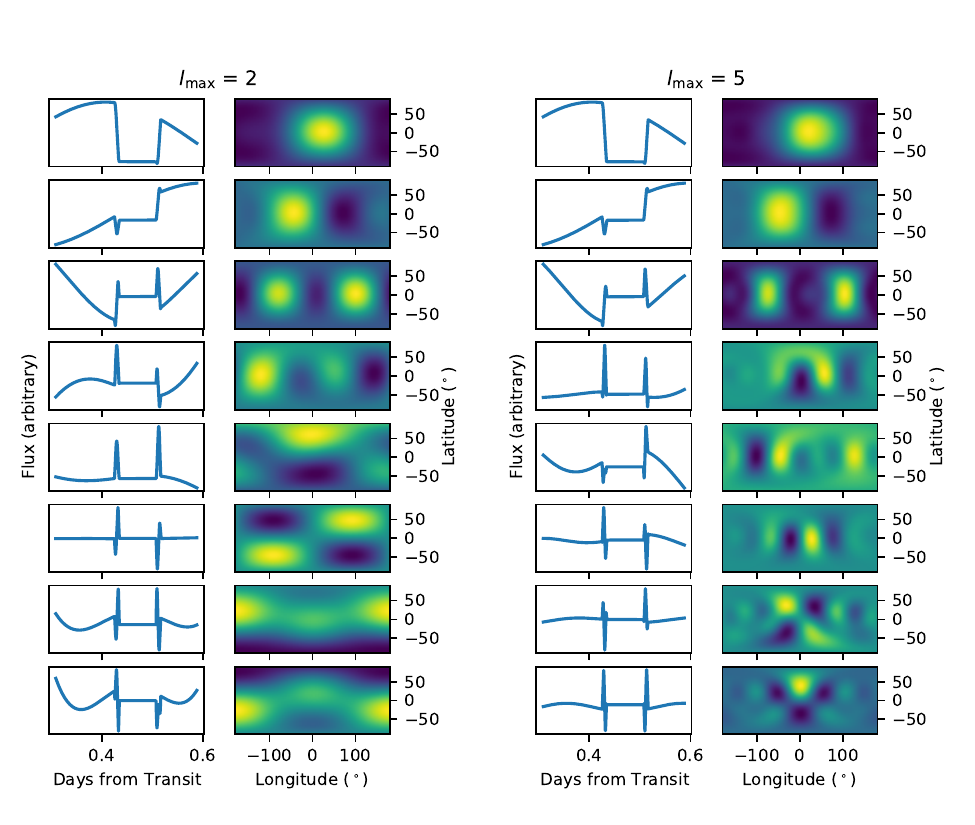}
    \caption{Two example sets of eigencurves and eigenmaps for $l_{\rm max} = 2$ (left) and $l_{\rm max} = 5$ (right) for the JWST NIRISS/SOSS observation in \cite{CoulombeEtal2023arxivWASP18b}. We show the top 8 eigencurves ranked top-to-bottom by their variance. The flux units on the eigencurves and eigenmaps are arbitrary, as they are scaled by the fitted parameters when modeling data.}
    \label{fig:maps}
\end{figure*}

In this work we use the ``eigenmapping'' method \citep{RauscherEtal2018ajMap, MansfieldEtal2020mnrasEigenspectraMapping, ChallenerRauscher2022ajThERESA} to model light curves using ThERESA \citep{ChallenerRauscher2022ajThERESA}.
Eigenmapping begins with a set of light curves generated from spherical harmonics of degree $\leq l_{\mathrm{max}}$.
These spherical harmonic light curves are then orthogonalized using principle component analysis (PCA) to generate ``eigencurves'' $E_i$\revision{.} 
The associated eigenvectors of these eigencurves directly correspond to spherical harmonic coefficients that can be used to \revision{calculate the ``eigenmaps'', which are the maps that, when integrated, generate the orthogonal eigencurves}.
The eigencurves are naturally ranked by their variance, or detectability (i.e., $E_1$ has the most variance, $E_2$ the second most, etc.).
Generally, $E_1$ corresponds to the amplitude of a planet's day-night contrast and $E_2$ adjusts for hotspot offsets.
Higher terms allow for smaller scale and latitudinal variation, but their precise structures are dependent on the planet's orbital parameters, making direct physical interpretation of higher-order terms difficult.
Figure \ref{fig:maps} shows two example sets of eigencurves $E_i$ for different \lmax\ and their corresponding eigenmaps, ranked top-to-bottom by their variance.

A light-curve observation is modeled as a linear sum of $N$ eigencurves, a uniform map component $Y_0^0$, and a correction term $s_{\mathrm{corr}}$ using the formula

\begin{equation}
\label{eq:eclmap}
    F_p(t) = \sum_{i=1}^N c_i E_i(t) + c_0 Y_0^0(t) + s_{\mathrm{corr}},
\end{equation}

\noindent
where $c_i$ (including $c_0$) are fitted parameters.
The associated map $Z$ is found by summing the corresponding eigenmaps and the uniform map, scaled in the same way by the fitted $c_i$ parameters.

There are two primary decisions to make in the eigenmapping method.
The first is the maximum degree of spherical harmonics ($l_{\mathrm{max}}$) to pass into the PCA. 
This effectively sets the maximum spatial resolution of variations in the map.
The second choice is how many eigencurve components $N$ to include in the fit.
A larger $N$ allows for greater model flexibility but there is a limit to the number of components supported by the data before one risks overfitting.
Throughout this work we use the notation LXNY to represent a model with \lmax\ = X and $N = $Y, i.e., L2N5 is a model with \lmax\ $= 2$ and $N = 5$.

Previous eclipse mapping analyses with ThERESA used the BIC to optimize $l_{\mathrm{max}}$ and $N$.
The BIC is convenient, as it includes a penalty for model complexity; additional model parameters must improve goodness-of-fit significantly to improve the BIC.
However, the BIC is a single number used to describe a model fit to an entire dataset, so it gives little insight into how the data are driving the inclusion (or rejection) of model parameters.
LOO-CV assigns a score to every data point for each model, enabling much more detailed model comparison (see \cite{WelbanksEtal2023ajLOOCV} for a discussion of these and other model selection metrics).

\section{Leave-one-out Cross-validation}
\label{sec:loocv}
LOO-CV is used to estimate the out of sample predictive accuracy of a model \citep{vehtari2017practical}. In LOO-CV, a model is fit to a dataset with one of the datapoint left out. The trained model is used to score the left-out data point by computing the expected log predicted density,
\begin{equation}
\text{elpd}_{\text{LOO}, i, \mathcal{M}} = \log p(\mathcal{D}_{i}|\mathcal{D}_{-i}, \mathcal{M}).
\label{eq:loo_score}
\end{equation}
Here, $\mathcal{M}$ is the model, $D_{i}$ is the $i$th datapoint and $\mathcal{D}_{-i}$ is the full dataset $\mathcal{D}$ with $\mathcal{D}_{i}$ left out. This quantity can by computed for each datapoint and can be summed over the dataset (with $N_d$ datapoints) to give an indication of overall model perforance,
\begin{equation}
    \text{elpd}_{\text{LOO},\mathcal{M}} = \sum_{i=1}^{N_d}\text{elpd}_{\text{LOO}, i, \mathcal{M}}.
    \label{eq:total_loo}
\end{equation}
Two competing models of the data ($\mathcal{M}_{1},\mathcal{M}_{2}$) can then be compared over the dataset with,
\begin{equation}
    \Delta\text{elpd}_{\text{LOO},\mathcal{M}_{1},\mathcal{M}_{2}} = \text{elpd}_{\text{LOO},\mathcal{M}_{1}} - \text{elpd}_{\text{LOO},\mathcal{M}_{2}}.
    \label{eq:loo_diff}
\end{equation}
Here a positive difference would indicate $\mathcal{M}_{1}$ has a better out of sample predictive performance than $\mathcal{M}_{2}$. Additionally, models can be compared at the data point level by examining the differences in individual scores in Equation (\ref{eq:loo_score}). Owing to the fact that we are estimating the out of sample predictive performance of a model with a finite data set, the standard error (SE) for the total LOO score and difference are given by,
\begin{equation}
    \text{SE} (\text{elpd}_{\text{LOO},\mathcal{M}}) = \sqrt{N_d\text{V}_{i=1}^{N_d}(\text{elpd}_{\text{LOO},i, \mathcal{M}})},
    \label{eq:se_loo}
\end{equation}
and,
\begin{align}
\begin{split}
    &\text{SE} (\Delta\text{elpd}_{\text{LOO},\mathcal{M}_{1},\mathcal{M}_{2}})
    \\&=\sqrt{N_d\text{V}_{i=1}^{N_d}(\text{elpd}_{\text{LOO},i, \mathcal{M}_{1}}-\text{elpd}_{\text{LOO}, i, \mathcal{M}_{2}})}.
\end{split}
    \label{eq:se_diff}
\end{align}
respectively. In Equations (\ref{eq:se_loo} and \ref{eq:se_diff}) V is the variance operator and $N_d$ is the number of data points. 

Naive computation of all terms in Equation (\ref{eq:loo_score}) would require $N_d$ refits of the model. For eclipse mapping, the data is the system (star and planet) flux vs.\ time, which for JWST/NIRISS is on the order of thousands per wavelength bin \cite[e.g., $N_d\approx2700$ for a single bin in][]{CoulombeEtal2023arxivWASP18b}. Fitting the models and obtaining posteriors samples with ThERESA typically takes $\approx50$ Central Processing Unit (CPU) hours. Due to the large number of data points combined with the large number of possible models considered in eclipse mapping, we turn the Pareto Smoothed Importance Sampling \citep[PSIS;][]{vehtari2015pareto} approximation to compute the terms in Equation (\ref{eq:loo_score}) as performed in 
\cite{vehtari2017practical}. 

In the PSIS approximation a model is fit to the entire data set and the resulting posterior samples are then re-weighted using importance sampling to approximate the effect of leaving out a data point to compute terms in Equation (\ref{eq:loo_score}). In addition to re-weighting the full posterior, PSIS also produces a diagnostic, Pareto k, which traces the accuracy of the approximation. Pareto k is used to determine whether to use the approximation or to perform a full refit which overall allows the accurate computation of all terms in Equation (\ref{eq:loo_score}) with significantly less than $N_d$ refits of the model. See \cite{vehtari2017practical} for a full explanation of the method and \cite{WelbanksEtal2023ajLOOCV} for a detailed account of the method applied to exoplanet spectroscopy. 

\section{Interpreting Eclipse Mapping with LOO-CV}

\label{sec:application}

Here we apply LOO-CV to the WASP-18b JWST eclipse observation presented in \cite{CoulombeEtal2023arxivWASP18b}, which is the highest-signal eclipse observation measured to date.
Previous eclipse-mapping efforts showed an ambiguity between two competing, distinct map models for this planet. Here we use LOO-CV to understand, and attempt to resolve, that degeneracy.

\subsection{WASP-18b Eclipse Mapping Thus Far}

\begin{figure}
    \centering
    \includegraphics{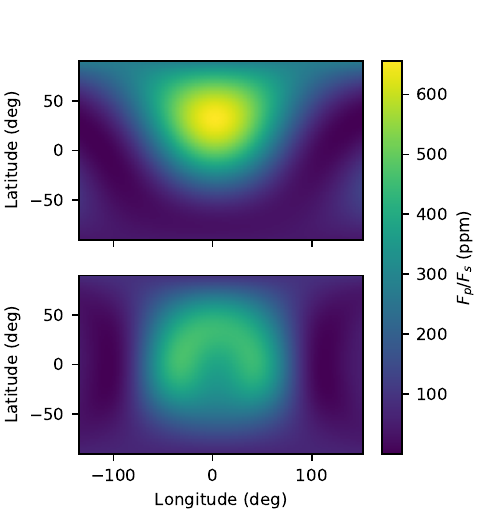}
    \caption{The two competing brightness map models of WASP-18b: L2N5 (top) and L5N5 (bottom). The L2N5 model has a peaked hotspot with a latitudinal offset, and the L5N5 model has a broader hotspot close to the equator. The longitudes have been restricted to those visible during the observation. Note that the uncertainties in the maps at longitudes on the nightside and high latitudes are significant, as these regions contribute minimal flux to the observed light curve.}
    \label{fig:w18maps}
\end{figure}

\cite{CoulombeEtal2023arxivWASP18b} presented the first JWST exoplanet eclipse maps of the ultra-hot Jupiter WASP-18b. 
They use precisely the mapping models discussed here, and find two equally-probably solutions: L2N5 and L5N5.
The eigencurve and eigenmap components of these fits are shown in the top five rows of Figure \ref{fig:maps} and the best-fitting maps are shown in Figure \ref{fig:w18maps}.
The L2N5 model has a peaked hotspot near $0^\circ$ longitude with a significant latitudinal offset.
The L5N5 model has a dimmer, broader hotspot ``plateau'' extending from $\approx -40 - 40^\circ$ longitude with less latitudinal variation. 

Aside from the large planet signal, WASP-18b is not an ideal eclipse-mapping target.
Its low impact parameter makes constraining latitudinal variation challenging, to the point where \cite{CoulombeEtal2023arxivWASP18b} average over the latitudinal variation and only make inferences from the longitudinal map.
Despite this challenge, they are able to make significant inferences from the eclispe map.
Both map models show little longitudinal offset of the brightest hemisphere of the planet and stark substellar-to-terminator temperature gradients, implying a drag, possibly of magnetic origin, that damps the super-rotating equatorial jet, potentially driving heat transfer over the poles \citep[e.g.,][]{BeltzEtal2021arxivWASP76bMagField}.

\subsection{Applying LOO-CV to the WASP-18b Eclipse}

\begin{figure}
    \centering
    \includegraphics[width=0.5\textwidth]{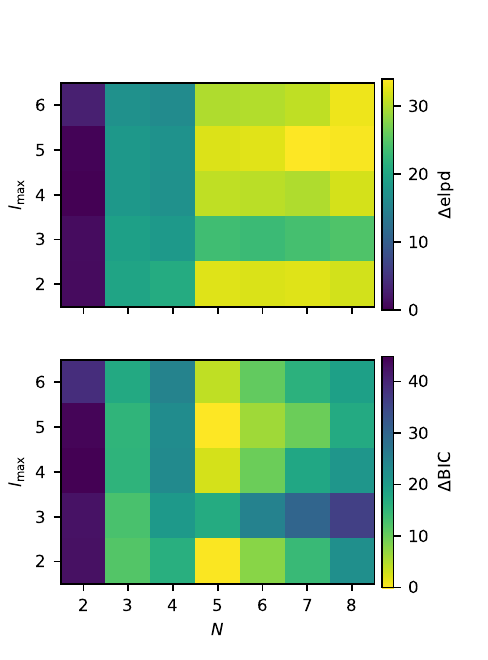}
    \caption{
    A comparison between models with different $l_{\rm max}$ and $N$ using different model selection criteria. For clarity we have left off $N=1$ and $l_{\rm max}=1$ fits, as they are all rejected with very high confidence. Note that higher $\text{elpd}_{\text{LOO},\mathcal{M}}$ and lower BIC are desired; this is reflected in the colorbars, where yellow indicates better performance. \textbf{Top:} $\text{elpd}_{\text{LOO},\mathcal{M}}$ for each combination of \lmax\ and $N$ relative to the lowest score in the grid. There is a clear jump in model preference when adding the fifth component, after which model improvement decreases. \textbf{Bottom:} BIC scores relative to the lowest score in the grid. The local minima at L2N5 and L5N5 are the two fits presented in \cite{CoulombeEtal2023arxivWASP18b} that are equally probable when considering the BIC alone.}
    \label{fig:loogrid}
\end{figure}

First, we investigate the total LOO-CV scores $\text{elpd}_{\text{LOO},\mathcal{M}}$ for each model fit for all models with $l_{\mathrm{max}} \leq 6$ and $N \leq 8$, which are the same models used in \cite{CoulombeEtal2023arxivWASP18b}.
The $\text{elpd}_{\text{LOO},\mathcal{M}}$, relative to the lowest score in the grid, are shown in Figure \ref{fig:loogrid}.
Note that, unlike the BIC, a higher $\text{elpd}_{\text{LOO},\mathcal{M}}$ indicates better performance, and that LOO-CV has no penalty for model complexity, so as components are added to the fit (moving right along a row) the $\text{elpd}_{\text{LOO},\mathcal{M}}$ increases until reaching a plateau.
In agreement with \cite{CoulombeEtal2023arxivWASP18b}, the highest scores are reached for \lmax\ $=2$ and \lmax\ $=5$, and plateau beyond $N = 5$.

\begin{figure*}
    \centering
    \includegraphics{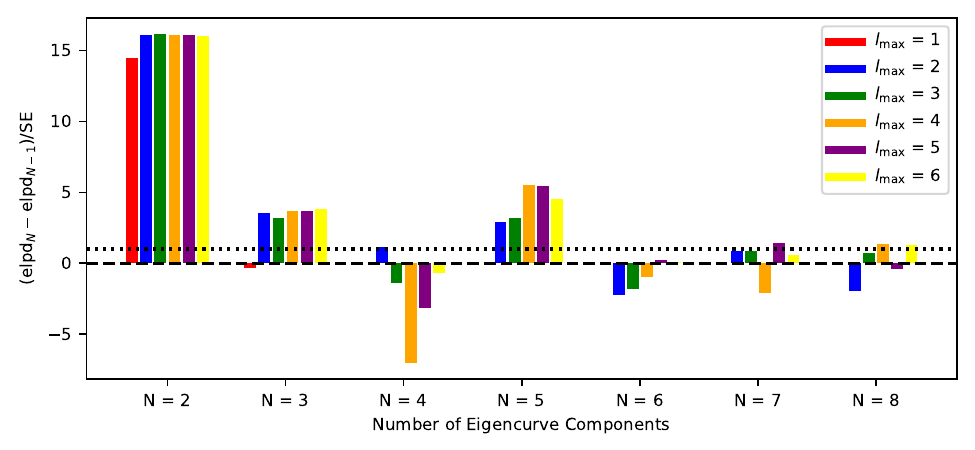}
    \caption{Comparison between models with $N$ and $N - 1$ model components with constant \lmax, in units of SE. A model is favored over the previous model if the addition of the new parameter increases $\text{elpd}_N$, and increases by more than 1 standard error are significant (i.e., the y-axis is $>1$, above the dotted line). Generally, the addition of the second, third, and fifth eigencurves are easily justified. However, the fourth eigencurve often results in a decrease in the predictive performance of the model.}
    \label{fig:se}
\end{figure*}

We can further inspect the out-of-sample predictive power of additional model components by examining the difference in $\text{elpd}_{\text{LOO},\mathcal{M}}$ between row-adjacent models in Figure \ref{fig:loogrid}. 
Figure \ref{fig:se} shows this difference between models with $N$ and $N - 1$ model components with constant \lmax, measured in units of standard error (SE). 
Positive and negative values indicate increased and decreased out-of-sample predictive performance of the added model component, respectively.
Values over one are interpreted as a significant improvement in predictive performance relative to the error.

First, we note that the out-of-sample predictive power of the $N = 2$ component is very significant ($\Delta\text{elpd}/\text{SE}>>1$) over a model with only the $N = 1$ component.
This shows that fitting the data requires the ability to adjust the location of the hotspot, or, in other words, we can reliably constrain the phase curve offset of the planet using just the eclipse and partial phase curve.

For most \lmax, adding the third eigencurve is also easily justified.
The exception is when \lmax\ $= 1$.
In this case, the input spherical harmonics are the three first-order harmonics: day-night contrast, east-west contrast, and north-south contrast. 
The day-night and east-west contrasts are measurable out of eclipse and, thus, have the highest variance, so these structures dominate $E_1$ and $E_2$.
The north-south contrast, which is only evident during eclipse ingress and egress, is left to $E_3$.
Adding $E_3$ to the fit does not improve the out-of-sample predictive power of the model, indicating that significant latitudinal variation (at least, of the sort present in first-degree harmonics) is either not present in the planet or its inclusion in the model is not justified by the data quality.
Either way, this latitudinal pattern cannot be detected, in agreement with previous findings \citep{CoulombeEtal2023arxivWASP18b}.

The fourth eigencurve leads to, at best, a marginal improvement in predictive power, and in most cases actually decreases the predictive power of the model. 
This could be interpreted as telling us that three eigencurve components is an appropriate choice for modeling these data.
However, for all \lmax, the fifth eigencurve leads to a signficant improvement in the out-of-sample predictive power of the model ($\Delta\text{elpd}/\text{SE}>1$).
By construction, the fourth eigencurve should be more detectable than the fifth eigencurve, but the $\text{elpd}_{\text{LOO},\mathcal{M}}$ show that the structures in the fourth eigencurve are not present in the planet at level that is evident in the data.
This suggests we should fit the data without $E_4$ but with $E_5$.

\begin{figure}
    \centering
    \includegraphics{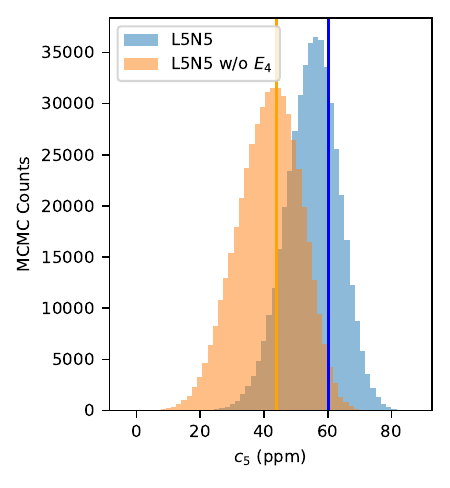}
    \caption{Posterior distribution of parameter $c_5$, the weight of the fifth eigencurve $E_5$, in models with and without the fourth eigencurve $E_4$ for $l_{\rm max} = 5$. The best-fit value of $c_5$ (vertical lines) changes depending on whether $E_4$ is in the model. While $E_4$ does not significantly improve the fit on its own, its presence allows for further flexibility in the value of $c_5$ within the positive-flux constraint, which does make a significant improvement to the fit. }
    \label{fig:no4thfit}
\end{figure}

We find that an L5N5 - $E_4$ model is preferred over L5N3 but not preferred over L5N5, with both BIC and LOO-CV. 
This somewhat unintuitive result is due to the requirement that the sum of the eigenmaps must be positive at locations of the planet which are visible during the observation. 
This constraint sets a limit on the sum of weights given to each eigencurve if they add constructively, eliminating regions of parameter space, effectively introducing correlations between model parameters (eigencurve/eigenmap weights).
If we introduce a new component to the model that adds destructively with a different component, this new component can increase the size of the allowed parameter space.
In this particular instance, while $E_4$ does not significantly improve the model fit to the data, including this term allows for further flexibility in $E_5$ within the flux positivity constraint, which does significantly improve the model fit (Figure \ref{fig:no4thfit}).
This insight into the parameter-space complexity introduced by the positive-flux constraint was, in part, made possible by the information gained from applying LOO-CV and highlights the importance of using an interpretable model-selection criterion.

Beyond $N = 5$, adding new components leads to decreased predictive performance of the model or only marginal ($<1$ SE) improvements.
Many of these lower-variance components contain latitudinal asymmetries (see Figure \ref{fig:maps}), which, as previously discussed, are difficult to detect with only one eclipse of WASP-18b, so non-detection of these components is expected.
This agrees with Figure \ref{fig:loogrid} and \cite{CoulombeEtal2023arxivWASP18b}, where they found that five eigencurves was the optimal choice for both competing planet models.

\subsection{Phase-resolved LOO-CV}

\begin{figure*}[t]
    \centering
    \includegraphics[width=\textwidth]{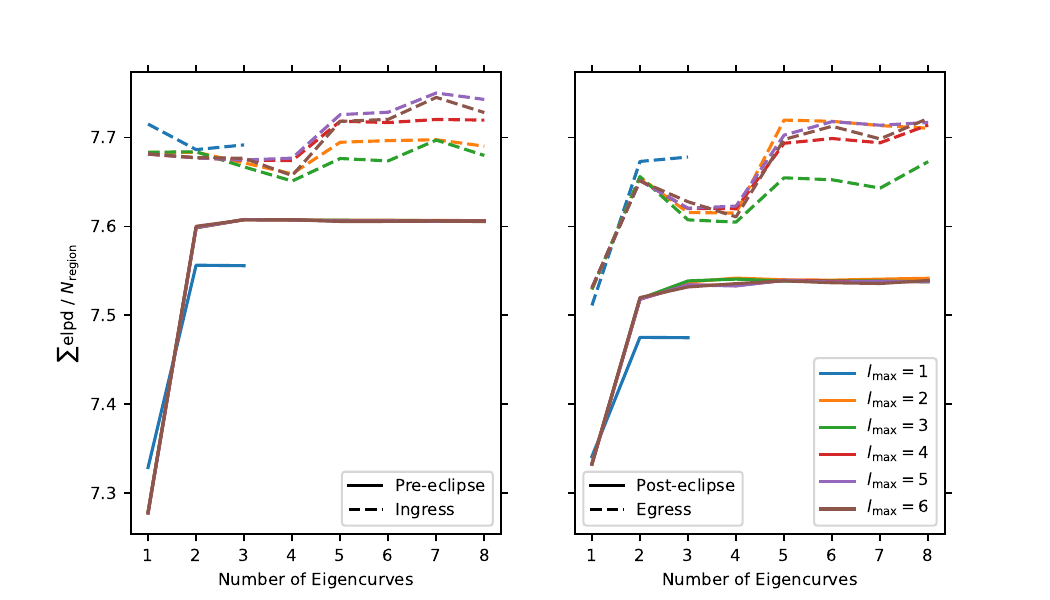}
    \caption{The total $\text{elpd}_{\text{LOO}, \mathcal{M}}$ over each region for 1 -- 8 eigencurves (1 -- 3 for $l_{\rm max} = 1$, as there are only three 1st-degree harmonics). Increases from left to right indicate that adding a given component to the model improves the model's out-of-sample predictive power in that region, and vice versa. Because adding eigencurves to the model does not affect the model during total eclipse (planet flux is zero), the change in model predictive power in that region is negligible and, thus, is not shown here.}
    \label{fig:looregions}
\end{figure*}

Until now, we have been considering the total $\text{elpd}_{\text{LOO}, \mathcal{M}}$ over the entire dataset.
However, we can sum the $\text{elpd}_{\text{LOO}, i, \mathcal{M}}$ over meaningful ranges of data to understand how additional eigencurves change model predictive power, just as we can sum over the whole data set to get a total $\text{elpd}_{\text{LOO}, \mathcal{M}}$ (Equation \ref{eq:total_loo}).
For example, by summing over pre-eclipse, ingress, total eclipse, egress, and post-eclipse we can tell which eigencurves improve the model predictive power over each region as a whole.
Figure \ref{fig:looregions} shows the per-data-point $\text{elpd}_{\text{LOO}, \mathcal{M}}$ for each of these regions as a function of $N$ and $l_{\rm max}$.
The regions are defined as follows: start of ingress $t_1 = 0.4258$, end of ingress $t_2 = 0.4348$, start of egress $t_3 = 0.5067$, and end of egress $t_4 = 0.5157$ days from transit.

First, we notice that in pre-eclipse and post-eclipse the model predictive power is primarily affected by the first three eigencurves, which agrees with our qualitative arguments above in regards to Figure \ref{fig:se}.
Increasing $l_{\rm max}$ beyond two leads to only slight improvements in elpd because the phase-curve variation outside eclipse is only sensitive to global-scale patterns which are represented by low-order harmonic modes.
The phase-curve model variation allowed by the higher-order terms is negligible relative to the uncertainties on the observation.
This quantitatively demonstrates the intuition applied in previous eclipse-mapping efforts \citep{ChallenerRauscher2022ajThERESA, CoulombeEtal2023arxivWASP18b}.

The story is more complex for ingress and egress.
For ingress, the fifth eigencurve is a marked improvement in the predictive power of the model for most $l_{\rm max}$, but the larger-scale (lower $N$) eigencurves are not significantly improving the model (the change in the sum of the $\text{elpd}$ is $\approx0$).
In fact, the fourth eigencurve decreases model performance.
Overall, the $l_{\rm max} = 5$ model performs the best, matching the findings of \cite{CoulombeEtal2023arxivWASP18b}. 
In egress, both the second and fifth eigencurves significantly improve the model, while the third is disfavored and the fourth does not change model performance. 
Here, the $l_{\rm max} = 2$ model performs the best.

Finally, for data during total eclipse (not shown), the $\text{elpd}_{\text{LOO},\mathcal{M}}$ does not change with number of eigencurves.
This is expected because the planet flux is zero during total eclipse, so any changes made to the planet model will not affect the light-curve model. 
The only model parameter which affects model flux during total eclipse is the stellar correction factor (Equation \ref{eq:eclmap}), which is always small (a few ppm) and consistent with zero as long as the data are accurately normalized \citep{ChallenerRauscher2022ajThERESA}.

By visualizing the $\text{elpd}_{\text{LOO},\mathcal{M}}$ in this way several things become clear:

\begin{enumerate}
    \item The second eigencurve (broadly, the hotspot offset) is primarily driven by the pre-eclipse and post-eclipse data. This matches intuition, as these regions contain the most data and are sensitive to large scale planet variations. The second eigencurve is also important to match the egress data.
    \item The third eigencurve (broadly, hotspot extent) is also primarily driven by out-of-eclipse data. While it represents a small improvement per data point, the total improvement is significant over all out-of-eclipse data.
    \item The fourth eigencurve leads to no improvement in the model in pre-eclipse, egress, and post-eclipse, and decreases model performance in ingress. Thus, the shape of ingress is driving the performance for the fourth eigencurve in Figure \ref{fig:se}.
    \item The fifth eigencurve also does not improve the model out-of-eclipse but makes substantial improvements to the model in both ingress and egress, confirming that the need for this eigencurve is driven entirely by the shape of the eclipse.
    \item Model performance in pre- and post-eclipse is unaffected by increasing $l_{\rm max}$ beyond 2. 
    \item The two competing planet models in \cite{CoulombeEtal2023arxivWASP18b}, L2N5 and L5N5, are driven by data in egress and ingress, respectively, with the preference for L5N5 in ingress being slightly greater than the preference for L2N5 in egress.
\end{enumerate}

\noindent
These insights are enabled by the per-data-point scoring capabilities of LOO-CV.

\begin{figure*}
    \centering
    \includegraphics[width=\textwidth]{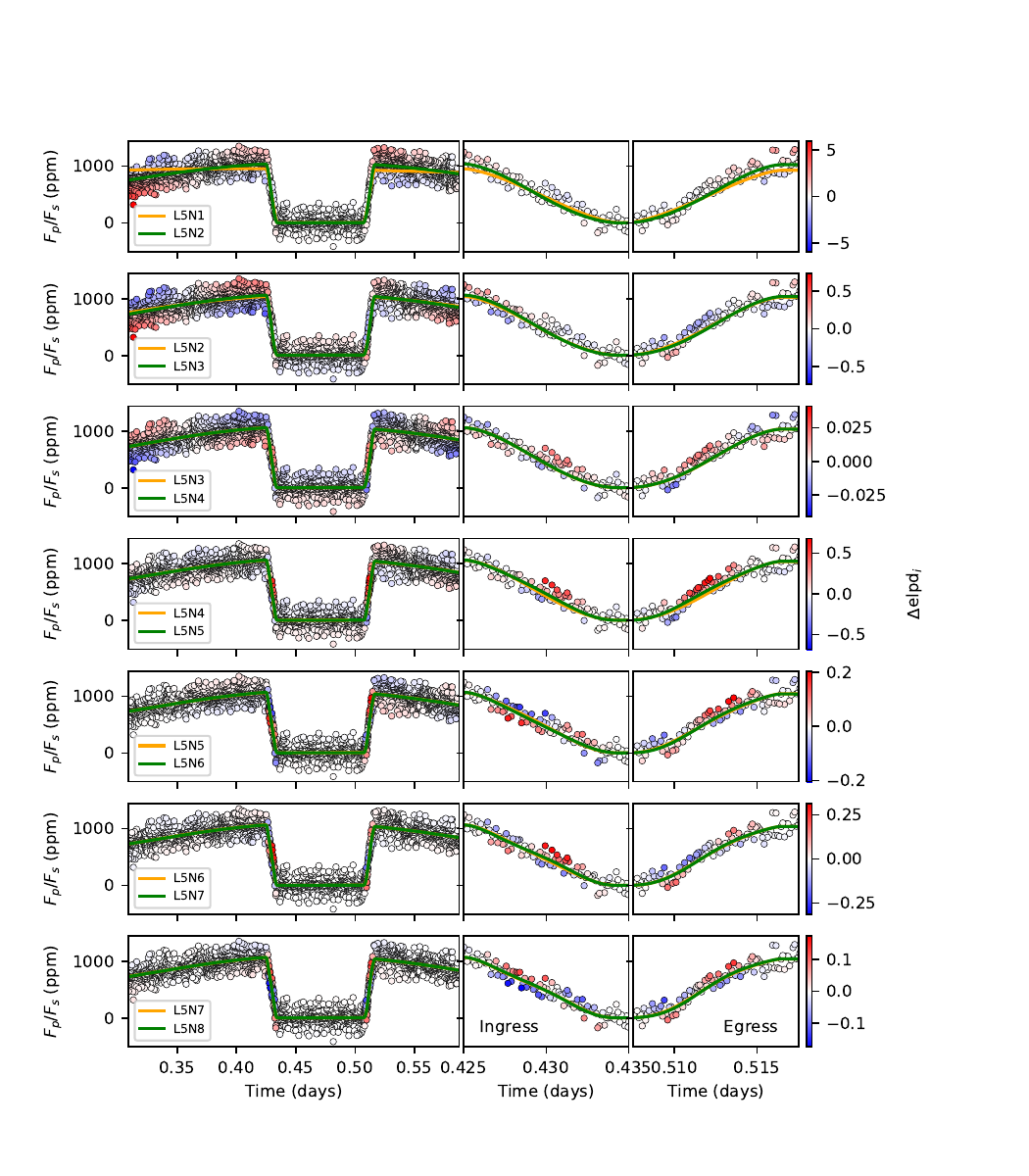}
    \caption{A comparison of models with model complexity increasing top-to-bottom. \revision{The left column shows the entire dataset, the middle column shows the eclipse ingress, and the right column shows the eclipse egress.} $\Delta \mathbf{elpd}_i$ is the difference in out-of-sample predictive power of the model with larger $N$ and the the model with smaller $N$. That is, positive values show where the additional parameters provides increased predictive performance, and vice versa. For visibility, the colorbars are on different scales.}
    \label{fig:loocomprow}
\end{figure*}

Uniquely, LOO-CV enables us to compare models data point by data point.
Figure \ref{fig:loocomprow} shows an example comparison between $l_{\rm max} = 5$ and $N \leq 8$ models overlaid on the data, which are color-coded by the difference in $\text{elpd}_\text{LOO}$. 
Examining a plot like this gives insight into how the data are driving the choice of $N$.
In the first two rows, which show the effects of adding $E_2$ and $E_3$ to the model, the $\text{elpd}_i$ differences are largest in the out-of-eclipse data, showing that these lower terms are primarily constrained by the phase curve variation.
The higher terms only show $\text{elpd}_i$ differences during ingress and egress, and, thus, their inclusion in the model is driven by the shape of the eclipse, rather than the phase curve.
Particularly, the $E_5$ term, which is statistically justified in the model from a BIC comparison \citep{CoulombeEtal2023arxivWASP18b}, is required due to signals only in eclipse ingress and egress; without the eclipse and an eclipse-mapping model, we would not be able to measure this brightness pattern.
While these conclusions are intuitive based on the eigencurves in Figure \ref{fig:maps}, LOO-CV gives us quantitative verification of this intuition.

\begin{figure*}
    \centering
    \includegraphics[width=\textwidth]{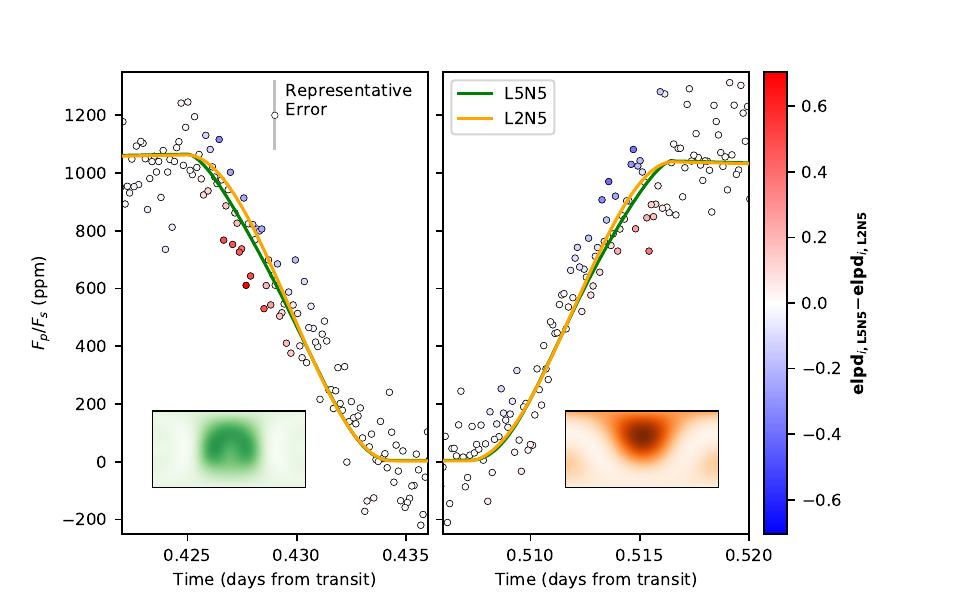}
    \caption{A comparison of $\text{elpd}_i$ between the two competing map models for WASP-18b, L5N5 and L2N5. The differences between the models are most significant in ingress (left) and egress (right). The insets show the flux map models (see Figure \ref{fig:w18maps}), in the same colors as their corresponding light-curve models. The L5N5 model is primarily preferred by data in the first half of ingress (red data points) while the L2N5 model is primarily preferred by data in the second half of egress (blue data points).}
    \label{fig:degeneracy}
\end{figure*}

We can make this comparison between the competing L5N5 and L2N5 models for WASP-18b to further understand which parts of the light curve are driving preference for one model over the other (Figure \ref{fig:degeneracy}).
Again we confirm that the L5N5 model is preferred for the ingress and the L2N5 model is preferred for the egress.
However, we now can see precisely which data points are driving the preference for each model.
Particularly, the L5N5 model performs better between $\approx0.426 - 0.430$ days and the L2N5 model performs better between $\approx0.513 - 0.516$ days.

By identifying the data points that are driving model preference, we can highlight regions where the data analysis may require further scrutiny.
For instance, we can look for correlations (e.g., unmodeled systematic or astrophysical signals) that might be affecting model preference and we can verify that our model selection is not significantly influenced by outliers.
For WASP-18b, the data in these regions are uncorrelated and there are no significant outliers, indicating that model preference is not driven by the map model attempting to correct for a signal unrelated to the planet's brightness map.
Similar verification may be important in future eclipse-mapping analyses where the data are less well-behaved.

\subsection{Recommendations for Eclipse-Mapping Analyses}

Here, we set forth a list of recommendations for how to leverage LOO-CV in eclipse-mapping analyses to better understand model selection and how the data inform model components:

\begin{enumerate}
    \item Verify that LOO-CV leads to the same model selection as the BIC (Figure \ref{fig:loogrid}). BIC minima should coincide with a plateau in the $\text{elpd}$, where adding further model components does not significantly improve the predictive power of the model. Lack of agreement may indicate that the BIC is not capturing the nuances of the data, so one should take a closer look at the $\text{elpd}_\text{LOO}$ over regions of interest and on a per-data-point basis.
    \item Inspect the change in $\text{elpd}$ against added model components in units of SE (Figure \ref{fig:se}). If any components that are included in the optimal model(s) are suggested to be unnecessary (i.e., $\Delta\text{elpd}/{\rm SE}$ is less than 1), verify that removing these components does not improve model selection criteria. Summing the $\text{elpd}$ over pre-eclipse, ingress, egress, and post-eclipse can pinpoint which region of the light curve is responsible for constraints, or lack thereof, on each model component (Figure \ref{fig:looregions}).
    \item For any models that are statistically indistinguishable, compare the $\text{elpd}_i$ to investigate which data points are driving the preference for each model (Figure \ref{fig:degeneracy}). Confirm that preference for none of the preferred models is driven by outliers or unmodeled systematic effects. If there are outliers or unmodeled systematic effects, consider removing the problematic data or further investigating the data reduction to eliminate correlations.
\end{enumerate}

\section{Conclusions}
\label{sec:conclusions}

Hot-Jupiter atmospheres are significantly non-uniform, with large day-to-night temperature contrasts and hotspots shifted away from the substellar point.
With JWST, these multidimensional properties of exoplanets are now measureable through eclipse mapping.
Furthermore, this multidimensionality can introduce biases into traditional 1D analyses.
Thus, eclipse mapping will be a crucial technique for understanding exoplanet atmospheres in the JWST era.

In this work, we used LOO-CV to investigate eclipse-mapping model selection and the ways in which the data inform model selection on a data-point-by-data-point basis. 
We demonstrated the use of LOO-CV with application to a JWST NIRISS/SOSS observation of WASP-18b, which has produced the first JWST eclipse maps of an exoplanet \citep{CoulombeEtal2023arxivWASP18b}.
We showed, quantitatively, that large scale features like day-night contrast and hotspot offset are primarily driven by the out-of-eclipse portions of the light curve, while smaller-scale features are constrained purely by the shape of the eclipse ingress and egress.
This confirms that the eclipse-mapping models are behaving as expected, and that with JWST we can leverage eclipse mapping to map planets in ways that are inaccessible to phase curves.

We also used LOO-CV to better understand the two competing map models of WASP-18b: a ``hotspot'' model and a ``plateau'' model.
A LOO-CV analysis shows that the plateau model is a better predictor of data in the early ingress while the hotspot model is a better predictor of data in the late egress.
Neither region of the light curve is afflicted by outliers or unmodeled signals, so neither model can be confidently rejected; determining the which map better matches the truth will likely require additional observations.
However, similar analyses could be extremely beneficial to future eclipse-mapping efforts.

Finally, we set forth a list of recommendations for the application of LOO-CV to eclipse-mapping analyses.
We suggest using LOO-CV to verify consistency in model selection criteria between LOO-CV and, for example, the BIC.
Then, we recommend applying LOO-CV to compare different models to check that model component constraints are behaving as expected; broad features should be constrained primarily by phase-curve variation and the inclusion of smaller-scale model components should be justified by the shape of the ingress and/or egress.
This approach is particularly useful for comparing models which are similar in preference, and can be used to investigate whether systematic effects are driving the preference for one model over another.

As our data quality improves, with JWST and future telescopes coming online, eclipse mapping will be a powerful tool to understand the multidimensionality of exoplanet atmospheres.
LOO-CV provides a detailed quantitative measure of how the data inform our models, significantly increasing confidence in eclipse-mapping results and the inferences on underlying physical processes of exoplanet atmospheres that follow.

\begin{acknowledgments}

  We thank contributors to SciPy, Matplotlib, Numpy, and the Python Programming Language. This work was supported by a grant from the Research Corporation for Science Advancement, through their Cottrell Scholar Award. Our work was conceptualized during the first EXOplanet Model Interrogation and New Techniques Sprint (EXOMINTS) held at Arizona State University in 2023. EXOMINTS and this project benefited from the 2022 Other Worlds Laboratory (OWL) Minigrant funded by the Heising-Simons Foundation. We thank Jonathan Fortney and the Other Worlds Laboratory for their support of this work.  L.W.~acknowledges support for this work provided by NASA through the NASA Hubble Fellowship grant \#HST-HF2-51496.001-A awarded by the Space Telescope Science Institute, which is operated by the Association of Universities for Research in Astronomy, Inc., for NASA, under contract NAS5-26555. This work was performed under the auspices of the U.S. Department of Energy by Lawrence Livermore National Laboratory under Contract DE-AC52-07NA27344. The document number is LLNL-JRNL-853759. \revision{We thank the anonymous referee for their comments which improved the quality of this manuscript.}
  \includegraphics[width=\linewidth,trim={0 0 0 0},clip]{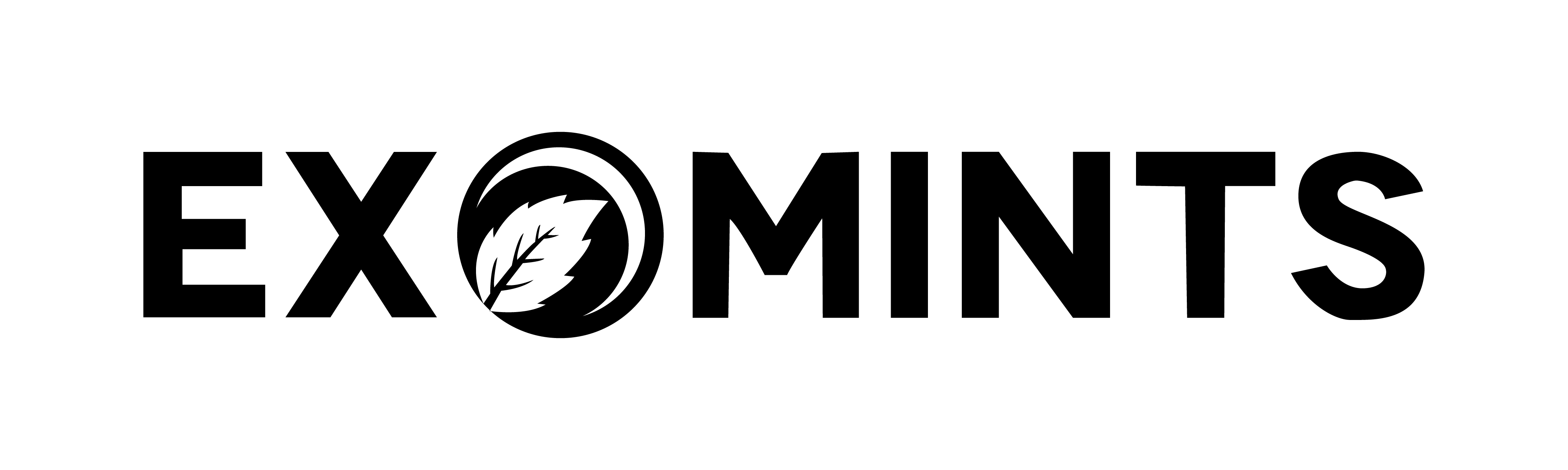}
\end{acknowledgments}

\software{NumPy \citep{HarrisEtal2020natNumPy}, Matplotlib
  \citep{Hunter2007cseMatplotlib}, SciPy
  \citep{VirtanenEtal2020natmSciPy}, Scikit-learn,
  \citep{PedregosaEtal2011jmlrScikitLearn}, starry
  \citep{LugerEtal2019ajStarry}, ThERESA \citep{ChallenerRauscher2022ajThERESA}, MC3 \citep{CubillosEtal2017ajRedNoise}}

\bibliography{loocv-mapping.bib}

\end{document}